\documentclass[notitlepage,twocolumn,prl,tightenlines,superscriptaddress,showpacs,floatfix]{revtex4-1} %nofootinbib %floatfix

\usepackage{amsmath,amssymb,amsfonts,latexsym}
\usepackage[colorlinks=true, linkcolor=blue, citecolor=blue]{hyperref}
\usepackage{bbm}
\usepackage[mathcal]{euscript}
\usepackage{graphicx}
\usepackage{epsfig}
\usepackage{color}
\usepackage{psfrag}
\usepackage{txfonts,pxfonts,wasysym,}
\usepackage{pifont}
\usepackage[caption = false]{subfig}
\newcommand{\be}{\begin{equation}}
\newcommand{\ee}{\end{equation}}
\newcommand{\ben}{\begin{equation*}}
\newcommand{\een}{\end{equation*}}
\newcommand{\ba}{\begin{eqnarray}}
\newcommand{\ea}{\end{eqnarray}}

\newcommand{\gul}{Gulliver UMR CNRS 7083, ESPCI Paris, PSL Research University, 10 rue Vauquelin, 75005 Paris, France}
\newcommand{\lisn}{LISN, UPR CNRS 3251, Universit\'e Paris-Saclay, 91405 Orsay, France}
\newcommand{\utem}{Departamento de Física, Facultad de Ciencias Naturales, Matemática y del Medio Ambiente, Universidad Tecnológica Metropolitana, Las Palmeras 3360, Ñuñoa 780-0003, Santiago, Chile.}

\begin{document}
\graphicspath{{./Figures/}}

\title{Laminar-Turbulent Patterns in Shear Flows : Evasion of Tipping,\\ Saddle-Loop Bifurcation and Log scaling of the Turbulent Fraction.}

\author{Pavan V. Kashyap}
\affiliation{\lisn}
\author{Juan F. Mar\'in}
\affiliation{\utem}
\author{Yohann Duguet}
\affiliation{\lisn}
\author{Olivier Dauchot}
\affiliation{\gul}

\date{\today}

\begin{abstract}
We analyze a one-dimensional two-scalar fields reaction–advection-diffusion model for the globally subcritical transition to turbulence. In this model, the homogeneous turbulent state is disconnected from the laminar one and disappears in a tipping catastrophe scenario. The model however exhibits a linear instability of the turbulent homogeneous state, mimicking the onset of the laminar-turbulent patterns observed in the transitional regime of wall shear flows. 
Numerically continuing the solutions obtained at large Reynolds numbers, we construct the Busse balloon associated with the multistability of the nonlinear solutions emerging from the instability. In the core of the balloon, the turbulent fluctuations, encoded into a multiplicative noise, select the pattern wavelength. On the lower Reynolds number side of the balloon, the pattern follows a cascade of destabilizations towards larger and larger, eventually infinite wavelengths.
In that limit, the periodic limit cycle associated with the spatial pattern hits the laminar fixed point, resulting in a saddle-loop global bifurcation and the emergence of solitary pulse solutions. This saddle-loop scenario predicts a logarithmic divergence of the wavelength, which captures experimental and numerical data in two representative shear flows.
\end{abstract}

\maketitle
The transitional regime to laminar flow in wall-bounded shear flows takes the form of coexisting laminar and turbulent domains~\cite{Manneville-2015,Tuckerman-2020}, resulting from a globally subcritical transition scenario~\cite{dauchot_local_1997}. When increasing the Reynolds number, $R$, the transition from the laminar flow to the regime of spatio-temporally intermittent turbulent spots follows a directed percolation scenario~\cite{pomeau_front_1986,lemoult_directed_2016,chantry_universal_2017,klotz2022phase}. 
Away from the transition, large aspect ratio plane Couette (pCf) and Taylor-Couette (TCf) flows, the flows sheared between two parallel planes, respectively two rotating coaxial cylinders, organize in the form of a regular laminar-turbulent pattern, as first experimentally reported in~\cite{Prigent-2002b,Prigent-2003a}. These patterns have now been reported experimentally or numerically in all wall-bounded shear flows~\cite{Tsukahara-2014,duguet_formation_2010,ishida2016transitional,kunii2019laminar,paranjape_onset_2019,Shimizu-2019,takeda_two-stage_2019,Kashyap-2020}, except for the circular Poiseuille flow (cPf)~\cite{Moxey-2010b}. In planar geometries they emerge, when decreasing R from homogeneous turbulent flow, with a wavelength $\lambda\simeq 50 h$, with $h$ the thickness of the direction of mean shear, and an orientation of approximately $25^{\circ}$ with respect to the mean flow~\cite{Prigent-2003a, Tsukahara-2014, Shimizu-2019, Kashyap-2020}. 

According to~\cite{Prigent-2002b}, the patterns result from a linear instability of the uniform turbulent flow, allowing for a weakly nonlinear Ginzburg–Landau description~\cite{Prigent-2003a}. This hypothesis was first reinforced by the study of the statistics of the spatial Fourier component of the pattern~\cite{Tuckerman-2008}. It recently found its confirmation with the study of the ensemble-averaged response of uniform turbulence to large-wavelength perturbations and the finding of a dispersion relation in very good agreement with the observed pattern~\cite{Kashyap-2022a}. This linear instability of the homogeneous turbulent flow also takes place in respectively two-fields~\cite{Manneville-2012} and six fields~\cite{Benavides-2023} models, obtained from the Navier-Stokes equations through specific closures.

%On the model side, a one-dimensional two-fields model, based on the Waleffe model for the streak–roll self-sustaining process~\cite{Waleffe-1997}, was shown to exhibit a Turing linear instability~\cite{Manneville-2012}. More recently, a two-dimensional six-fields model, obtained from the Navier-Stokes equations through specific closures, reproduces the observed phenomenology and predicts the right wavelength and orientation of the pattern from the linear stability analysis of the homogeneous turbulent state~\cite{Benavides-2023}.
%
\begin{figure}[t!]
    \centering
    \includegraphics[width=\columnwidth]{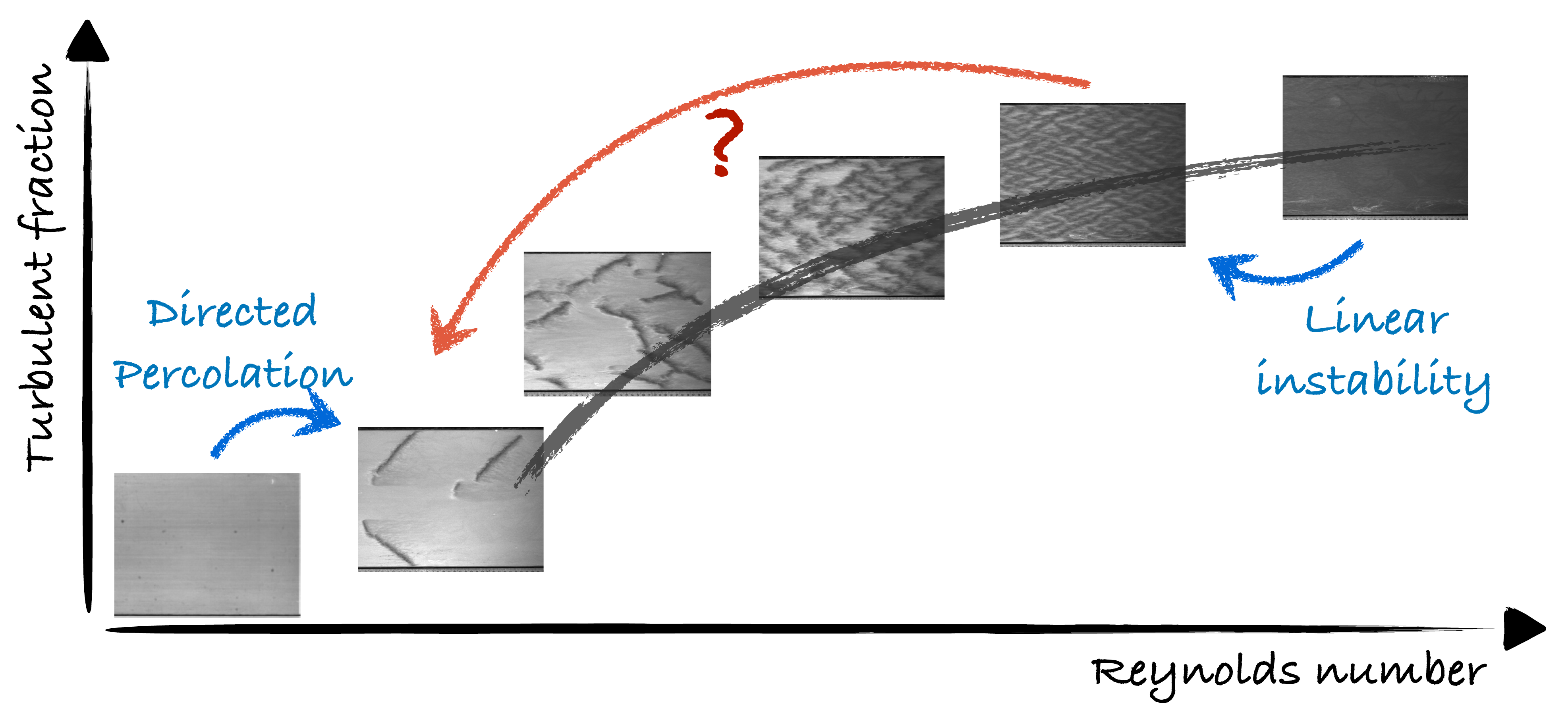}
    \vspace{-5mm}
    \caption{{\bf The subcritical transition to turbulence in plane shear flows:} Decreasing the Reynolds number from the homogeneous featureless turbulence, a laminar-turbulent pattern with a distinct wave vector emerges. Our central goal is to understand by which phase space mechanism the pattern evolves towards isolated structure before only the laminar flow eventually subsists. The experimental snapshots~\cite{paranjape_onset_2019}, illustrate this scenario in the case of the plane Poiseuille flow (pPf).}
    \label{fig:1}
    \vspace{-7mm}
\end{figure}

Here we tackle the remaining central issue illustrated in Fig.~\ref{fig:1}, namely how the strongly nonlinear evolution of the pattern leads, in a characteristic tipping evasion scenario~\cite{rietkerk2021evasion}, to the spatio-temporally intermittent regime when decreasing $R$. Starting with the simplest two-fields model~\cite{Manneville-2012}, to which advection and noise are added, we reveal rich unexplored dynamics. More specifically, we show that (i) many stable patterns coexist with wave vectors delimited by the boundaries of the so-called Busse balloon~\cite{busse_non-linear_1978}, (ii) a specific wavelength is selected when adding a multiplicative noise that mimics the turbulent fluctuations, (iii) a cascade of instabilities lead to a dramatic increase of the wavelength on the lower $R$ side of the balloon, eventually reaching infinite wavelength, (iv) in this limit, the periodic limit cycle associated with the spatial pattern hits the laminar fixed point into a saddle-loop global bifurcation. 
According to this scenario, the pattern evolves from a regular harmonic pattern to a non-harmonic one, with a logarithmic divergence of its wavelength, while the turbulent regions remain essentially of constant size. The prediction compares well with existing data from the literature, as long as the pattern is well formed, i.e. before entering the spatio-temporal intermittent regime obeying the directed percolation scenario. 

The Waleffe model~\cite{waleffe_self-sustaining_1997,Dauchot-2000a} was decisive in understanding the local self-sustaining process along which streamwise vortices ($V$) draw the mean flow ($M$) from high to low-velocity regions leading to streamwise velocity fluctuations called streaks ($U$). The latter undergo an inflectional instability leading to modulations of amplitude ($W$), feeding back the vortices and thereby sustaining turbulence. An important step forward was to propose a spatially extended version of the model including diffusive coupling~\cite{Manneville-2012}. Assuming fast dynamics for the fields $U$ and $V$, the model was reduced to a two-field model, which exhibits a Turing linear instability of the homogeneous turbulent flow, when decreasing $R$. Here we further expand on this first success by adding advection and a multiplicative noise to account for the fluctuating nature of the turbulent field $W$. The model we study in one dimension then reads: 
\begin{subequations}
\label{eq:PDE}    
\begin{align}
\frac{\partial M}{\partial t} + M \frac{\partial M}{\partial x} &= f(M,W) + D_M \frac{\partial^2 M}{\partial x^2}, \label{eqm}\\
\frac{\partial W}{\partial t} + M \frac{\partial W}{\partial x} &= g(M,W) + D_W \frac{\partial^2 W}{\partial x^2} + \sigma W \eta \label{eqw},
\end{align}
\end{subequations}
where $D_M,D_W$ are the diffusivities of the fields $M,W$, $\sigma$ is the amplitude of the noise and $\eta$ is a Gaussian random field with zero mean and $\left<\eta(x,t)\eta(x',t')\right>=\delta(t-t')\delta(x-x')$ (see Supp. Mat. sec.~I-A,B for more details on the model ). 
The functions $f(M,W)$ and $g(M,W)$ are given by:
\begin{subequations}
\label{eq:flow}    
\begin{align}
f(M,\hspace{-.01cm}W)\! &=\! -\alpha_{_{M}} (M-M_0) - \beta_{_{M}} (M - \bar{M} ) W^4 + \sigma_{_{M}} W^2, \label{feq}\\
g(M,\hspace{-.01cm}W)\! &=\! -\alpha_{_{W}} W + \left[\beta_{_{W}} (M - \bar{M})- \gamma_{_{W}}\right] W^3 - \sigma_{_{M}} M W ,\label{geq} 
\end{align}
\end{subequations}
where the dependence on $R$ is encoded in the coefficients (see Supp. Mat. sec.~I-B). The absorbing nature of the laminar state $(M=M_0, W=W_0=0)$ emerges from the form of $f$ and $g$ and is respected by the choice of a multiplicative noise. 

\begin{figure}[t!]
    \centering
    \includegraphics[width=\columnwidth]{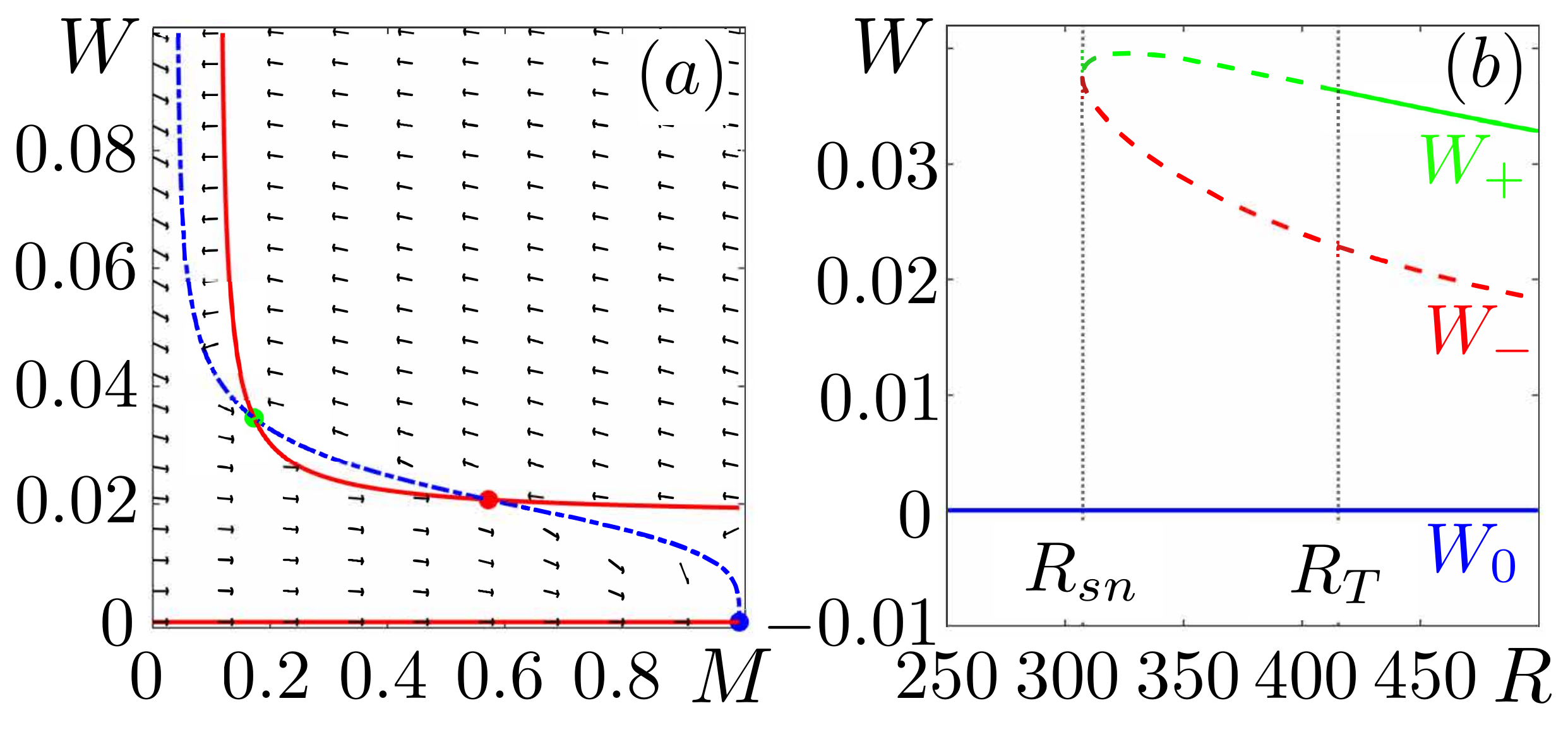}
    \vspace{-5mm}
    \caption{{\bf Homogeneous solutions:} \textbf{(a)} Vector field $(f,g)$ (black arrows) and nullclines $f(M,W)=0$ (blue), and $g(M,W)=0$ (red)  for $R=450>R_{sn}=310$. The green, red and blue dots respectively indicate the upper $(M_{+},W_{+})$, lower $(M_{-},W_{-})$, and laminar $(M_0, W_0)$ solutions. \textbf{(b)} Bifurcation diagram for the homogeneous field $W$ as a function of $R$. The laminar state (blue) is linearly stable for all $R$. The lower $W_{-}$, and upper $W_{+}$ branches emerge from a saddle-node bifurcation at $R=R_{sn}=310$. $W_{-}$ is linearly unstable for all $R>R_{sn}$. $W_{+}$ is linearly stable at large $R$ and becomes unstable via a Turing instability for $R<R_T=415$.}
    \label{fig:2}
    \vspace{-5mm}
\end{figure}
Figure~\ref{fig:2} displays the vector field $(f,g)$, together with the nullclines $f(M, W)=0$ and $g(M, W)=0$, the intersection of which provide the spatially homogeneous fixed points. The laminar solution $(M=M_0, W=W_0=0)$ is linearly stable for all $R$. For $R\ge R_{sn}$, two extra solutions $(M=M_{\pm},W=W_{\pm})$ emerge from a saddle node bifurcation.
The lower branch $(M_{-}, W_{-})$ is linearly unstable for all $R>R_{sn}$
The parameter values are 
%set $(\tilde\alpha_M=12.35, \tilde\beta_M=0.0013, \tilde\sigma_M=0.062, \tilde\alpha_W=1.426, \tilde\beta_W=9.6773\times10^{-4}, \tilde\sigma_W=0.68, \tilde M=18.3777, D_M=10, D_W=2$), 
such that the upper branch $(M_{+}, W_{+})$ is linearly stable with respect to homogeneous perturbations for $R>R_{sn}=310$, yet destabilizes via a Turing instability, when decreasing $R$ below $R_T=415$ (see also~\cite{Dauchot-2000a, Manneville-2012} and Supp. Mat. sec.~I.C).

\begin{figure}[t!]
    \centering
    \includegraphics[width=\columnwidth]{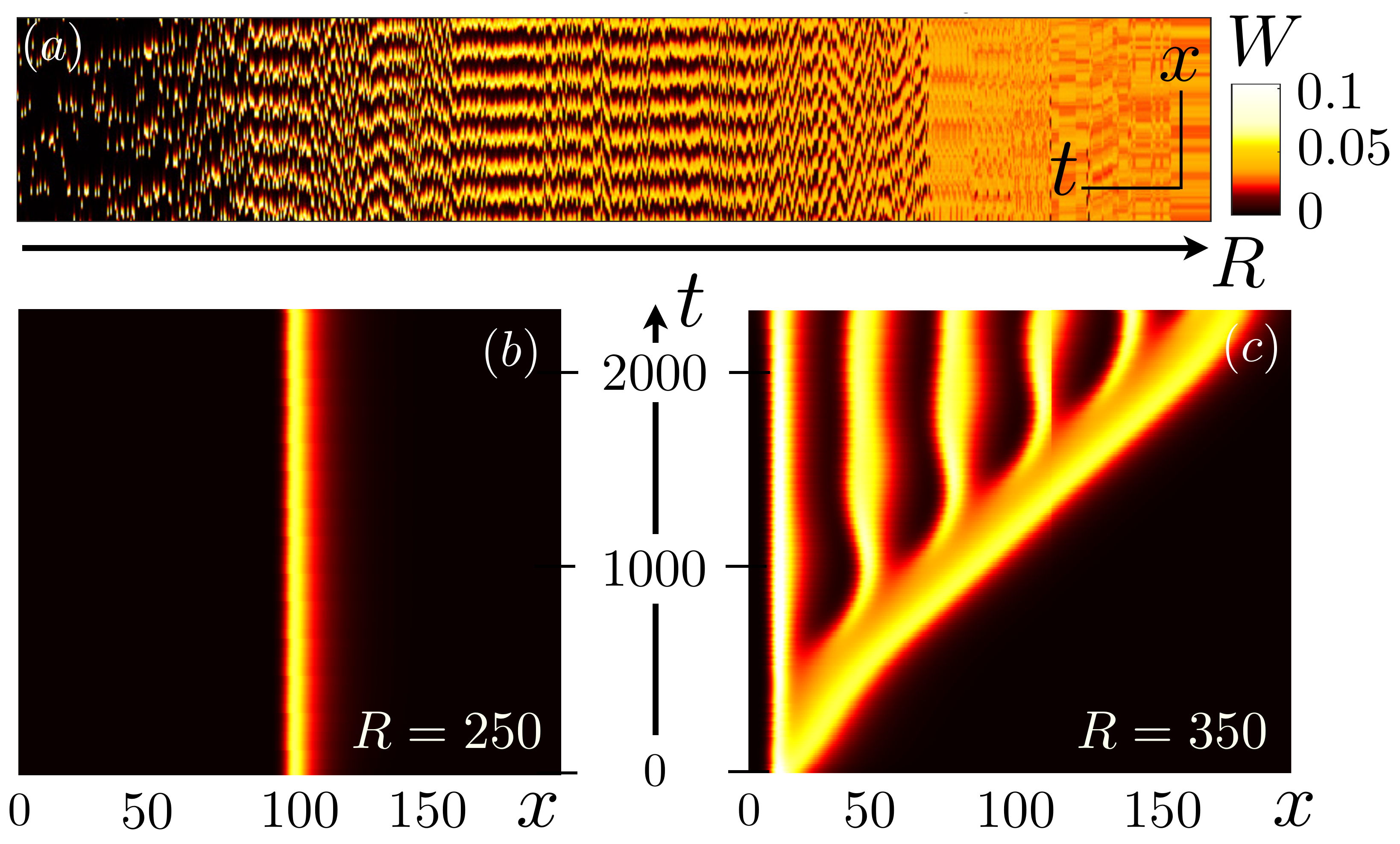}
    \vspace{-5mm}
    \caption{{\bf Phenomenology of the model:} Space-time plot of $W(x,t)$ in the advected frame (a) during the slow annealing of a homogeneous turbulent solution, decreasing $R$ from 500 to 200; (b) for a localized pulse with $R=250<R_{sn}$; (c) for a growing pattern with $R=350>R_{sn}$.}
    \label{fig:3}
\vspace{-5mm}    
\end{figure}

The simulation of Eqs.~(\ref{eq:PDE}) with periodic boundary conditions in a domain of size $L=1000$ (see Supp. Mat. sec.~II-A) display the main features of the transitional regime in plane shear flows. When slowly annealing $R$ from the homogeneous turbulent state (Fig.~\ref{fig:3}-a), a periodic pattern emerges, the wavelength of which increases with decreasing $R$. For low enough values of $R$, the pattern is replaced by a disordered juxtaposition of isolated pulses, which eventually vanish when further decreasing $R$ below $R_g\simeq 200$. 
For $R<R_{sn}$, the model exhibits an excitable-like behaviour: when initialized with a large enough amplitude Gaussian profile of $W$, a localized pulse nucleates (Fig.~\ref{fig:3}-b). Single pulses, as well as superposition of multiple pulses, are stable. 
The amplitude of the initial condition necessary to trigger a pulse increases with decreasing $R$. 
Nucleating two pulses one after the other either leads to a fusion of the two pulses when they are too close or to two distinct pulses, with the downstream one being pushed away from the upstream one. 
For $R>R_{sn}$, the initial Gaussian profile expands downstream, increasing the width between a laminar-turbulent front and a turbulent-laminar one. The turbulent state between the two fronts modulates, leading to a well-formed pattern (Fig.~\ref{fig:3}-c), as also seen in~\cite{alavyoon_turbulent_1986, duguet_formation_2010}.

\begin{figure}[t!]
    \centering
    \includegraphics[width=\columnwidth]{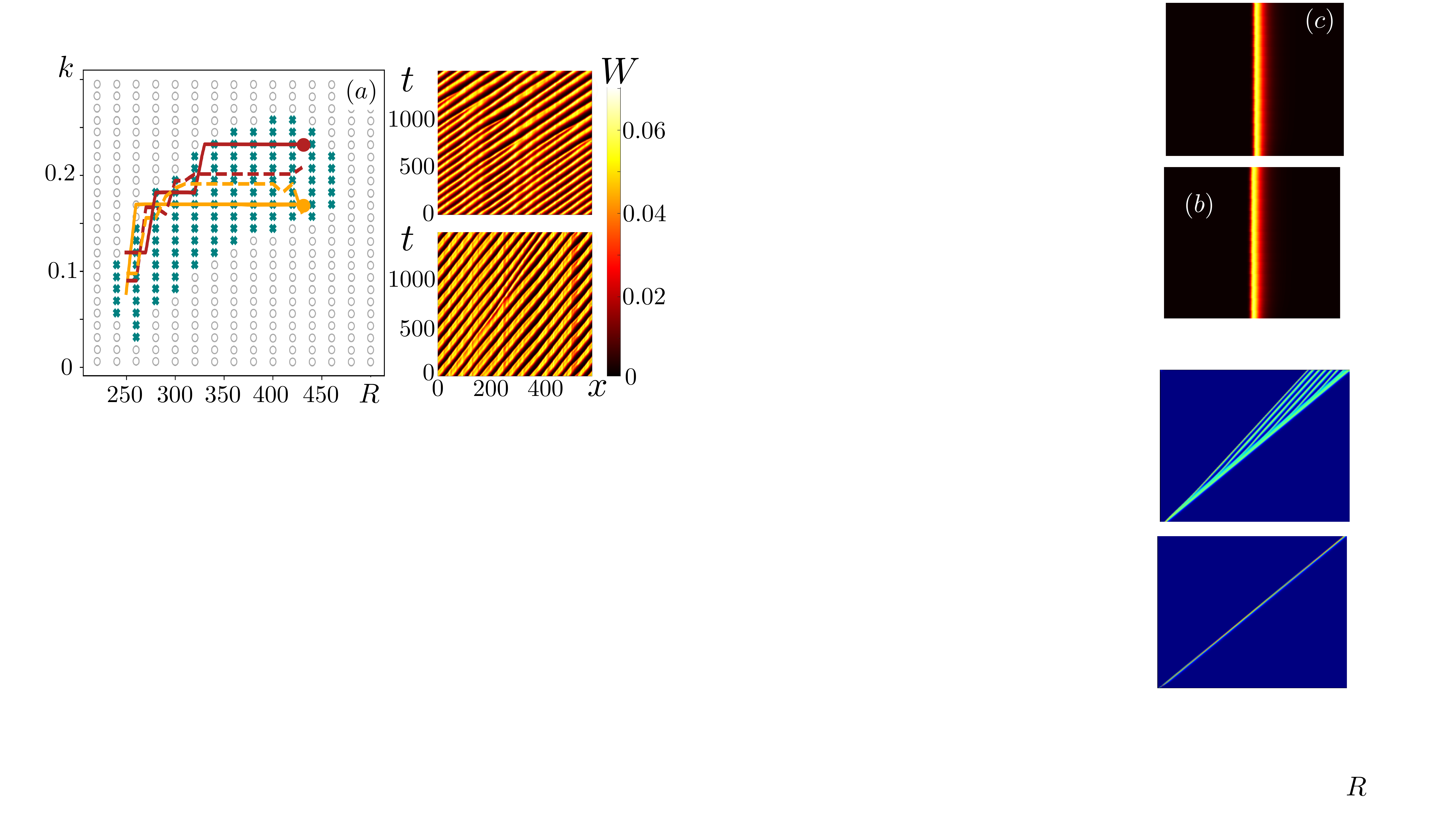}
    \caption{{\bf Selection of the pattern by noise :} (a) Busse balloon showing the stable patterns for the deterministic dynamics in the $(k, R)$ space, with annealing trajectories followed for two initial conditions (brow and yellow dots) without (continous lines) and with (dashed lines) noise; (b,c) Space-time plot of $W(x,t)$ showing the transient dynamics from an initial condition with wavelength smaller (b), respectively larger (c), than the selected one.}
    \label{fig:4}
\vspace{-5mm}    
\end{figure}

The pattern wavelength observed when applying the above annealing procedure is selected by noise. To show this, we first run deterministic simulations ($\sigma = 0$) with initial conditions in the form of a harmonically modulated turbulent state. We identify the stable patterns as those which keep the same wave number $k=2\pi/\lambda$ for very long times ($T>100\,000$). The region in the $R-k$ space encompassing all the stable patterns is called the Busse balloon~\citep{busse_non-linear_1978}. It extends for $R$ up to $460> R_T$, retaining the subcriticality of the Turing instability reported in~\cite{Manneville-2012}, and down to $\simeq 240$, far below $R_{sn}$, in a characteristic evasion from the tipping point catastrophe~\cite{rietkerk2021evasion}. 
When initiating the noiseless simulation with a well-formed pattern and decreasing the Reynolds number, the wavelength remains constant until the pattern hits the left edge of the Busse balloon (continuous line trajectories on Fig.~\ref{fig:4}-a). Conversely, in the presence of noise, patterns initiated with two different wave numbers converge towards a narrow range of wavelength, which is also the one obtained when annealing from the homogeneous turbulent state (dashed line trajectories Fig.~\ref{fig:4}-a and spatio-temporal diagrams Fig.~\ref{fig:4}-b,c).

\begin{figure}[t!]
    \centering
    \includegraphics[width=\columnwidth]{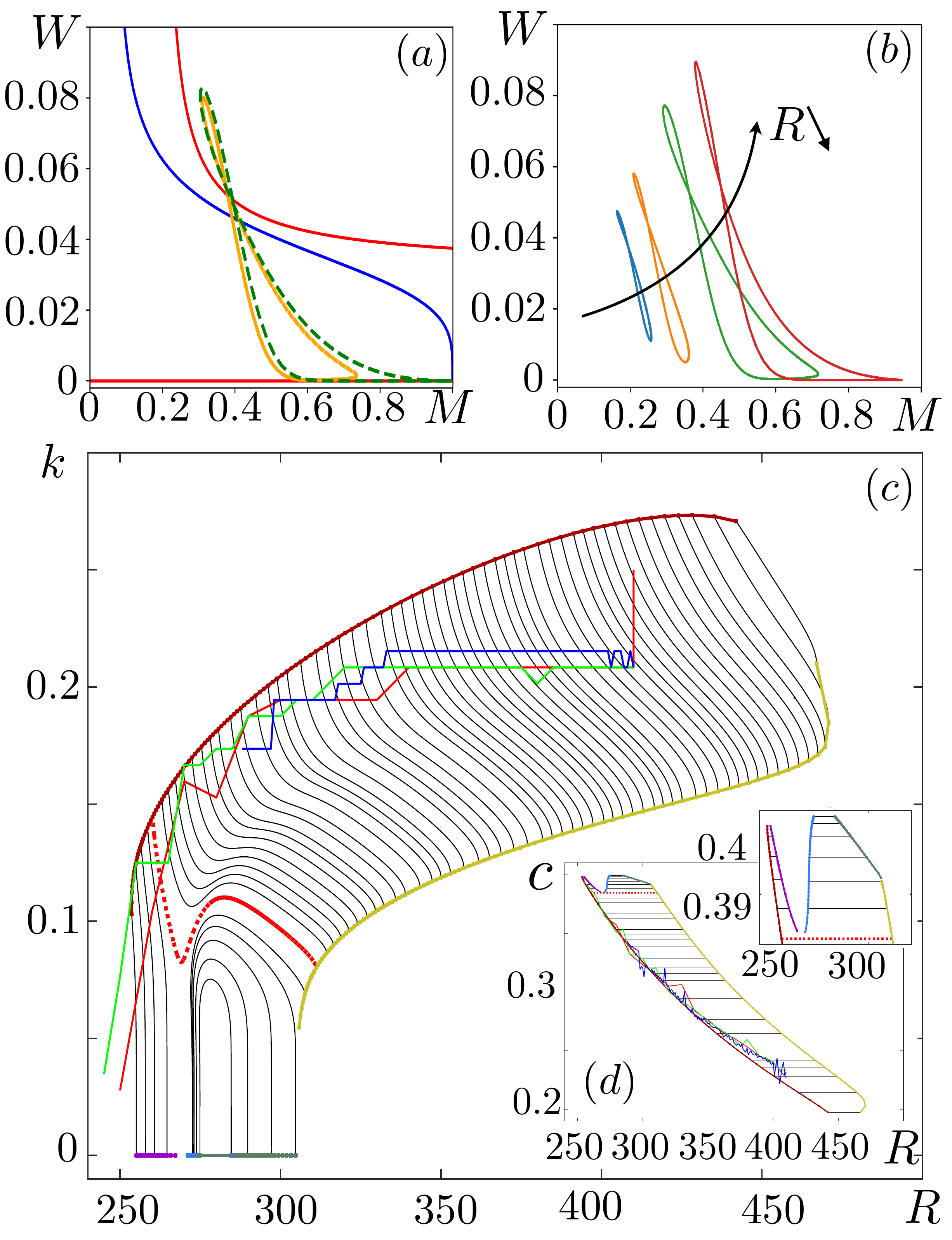}
    \vspace{-3mm}
    \caption{{\bf Saddle-loop bifurcation: }(a) Parametric plot in the $(M, W)$ space of a pattern (continuous yellow line) and a pulse (dashed green line) solution of Eqs.~(\ref{eq:PDE}), coexisting at $R=250$; (the blue and red lines are the nullclines of $f$ and $g$); (b) Same representation for periodic pattern solutions of Eqs.~(\ref{eq:PDE}), obtained for $R=220, 250, 350, 430$; (c-d) Busse balloon, in $(k, R)$ and $(c, R)$ space, as obtained from numerical continuation of the periodic solutions of Eq.~(\ref{eq:ODE}); in blue green and red are the path, followed by the solutions of Eqs.~(\ref{eq:PDE}), following three independent annealing procedures; (inset: zoom on the large $c$, small $k$, solutions).}
    \label{fig:5}
\vspace{-5mm}    
\end{figure}

When $R\lesssim R_{sn}$, large wavelength patterns coexist with localized pulses in phase space. The connection between these two families of solutions is better described in the system of coordinate $z=x-ct$, where $c$ is the advection speed of the travelling solutions. One thereby obtains a 4-dimensional dynamical system: 
\begin{subequations}
\label{eq:ODE}
\begin{align}
\dot M &= P, \\
D_{_M} \dot P &= (M-c) P - f(M,W), \\
\dot W &= Q, \\
D_{_W} \dot Q &= (M-c) Q - g(M,W), 
\end{align}
\end{subequations}
with $\dot A \equiv dA/dz$. The fixed points of this dynamical system are $(M=M_{0,\pm},P=0,W=W_{0,\pm},Q=0)$ and are all saddles with two positive eigenvalues. The pattern and the pulse solutions of Eqs.~(\ref{eq:PDE}) respectively map onto periodic orbits around the turbulent fixed point and homoclinic orbits of the laminar one (see Fig.~\ref{fig:5}-a).
When reducing $R$, the periodic orbits grow around the unstable homogeneous turbulent state (see Fig.~\ref{fig:5}-b), distort and approach the laminar saddle. The central question is whether a global saddle-loop bifurcation takes place, with the periodic orbits turning into homoclinic ones~\cite{strogatz_nonlinear_2000,doelman2018destabilization}. If such a scenario were to take place, it would strongly constrain the wavelength dependence on $R$ in the vicinity of the global bifurcation~\citep{gaspard_measurement_1990}. This question can however not be answered using simulations with periodic boundary conditions, which formally only produce periodic solutions.

We thus turn to numerical continuation methods to follow the periodic orbits of Eqs.~(\ref{eq:ODE}) and their stability~\cite{sherratt2012numerical} (see Sup. Mat. sec.~II-B). A pattern solution at large $R\simeq 450$, which advection speed $c\simeq 0.2$ is measured, is seeded as a solution for Eqs.~(\ref{eq:ODE}). We then perform arc-length continuation, both increasing and decreasing $R$, keeping $c$ fixed (black lines of Fig.~\ref{fig:5}-c,d), until a secondary bifurcation takes place (red and yellow bounds of Fig.~\ref{fig:5},c-d). Starting from a stable solution, close to the lower $R$ instability, we then suggest a small increase in $c$ and iterate the procedure until the full Busse balloon, with a highly non-trivial shape at low $k$, is obtained.
In particular, we find two families of solutions with arbitrarily large periods located in the two "legs", resp. "tips", in the $(k, R)$, resp. $(c, R)$, representation of the Busse balloon. Those solutions eventually connect to the homoclinic orbits starting from the laminar fixed point and therefore confirm the global saddle-loop bifurcation scenario.

Fig.~\ref{fig:5}-(c,d) also displays the trajectory followed by the pattern solution in the $(R,k)$ and $(R,c)$ parameter spaces during three independent annealing processes as the one reported in Fig.~\ref{fig:3}-(a). The pattern selected by the noise at large $R$ moves across the Busse balloon, maintaining an essentially constant wavelength until it reaches its left boundary and follows it down to larger and larger wavelength. The growth of the wavelength is then controlled by the saddle-loop bifurcation and, therefore, obey the following logarithmic scaling~\cite{gaspard_measurement_1990}:
\begin{equation}
\label{eq:logdiv}
    \lambda = -a \log \epsilon + b,
\end{equation}
with $a,b>0$ and where $\epsilon=(R-R_{sl})/R_{sl}$, with $R_{sl}\simeq 250$ the critical value of $R$, where the saddle-loop bifurcation takes place.

\begin{figure}[t!]
    \centering
    \vspace{-1mm}
    \includegraphics[width=\columnwidth]{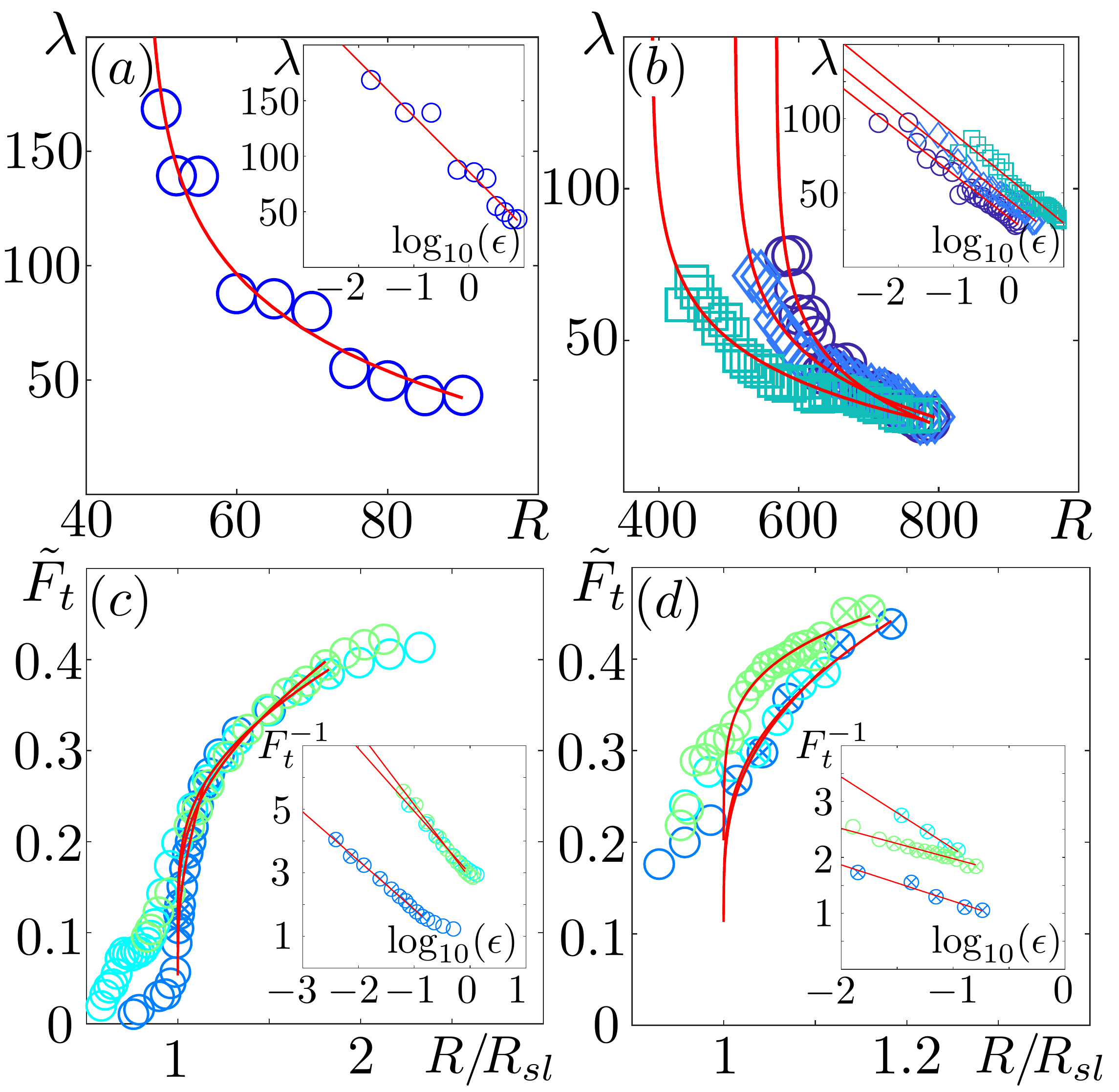}
    \caption{{\bf Log scaling of the wavelength and turbulent fraction in real flows} (a-b) Wavelength and (c-d) turbulent fraction reported in (a) simulations of pPf~\cite{Kashyap-2020}; (b) experiment with TCf for three angular speeds of the outer cylinder~\cite{prigent_spirale_2001,Prigent-2003a}; (c) three independent simulations of the pPf~\cite{Tsukahara-2014,Shimizu-2019,Kashyap-2020}; (d) simulation of the plane Couette flow (pCf)~\cite{duguet_formation_2010} (dark blue), experiment with large aspect ratio pCf~\cite{prigent_spirale_2001} (light blue), experiment with tCf, with both cylinders rotating at opposite speed~~\cite{prigent_spirale_2001} (light green). Fits are all obtained according to the logarithmic scaling~(\ref{eq:logdiv}) (see Supp. Mat. sec.~III)}
    \label{fig:6}
\vspace{-5mm}
\end{figure}

In real flows, the saddle-loop bifurcation scenario certainly does not strictly take place, first because of finite size effects and second because other mechanisms such as longitudinal band splitting~\cite{Shimizu-2019} may take over. Nevertheless, the proximity of a global bifurcation should control the behaviour of the wavelength in a critical regime for $R\gtrsim R_{sl}$. We conclude this work by testing this hypothesis with existing experimental and numerical measurements of both the pattern wavelength and the turbulent fraction in plane Poiseuille flow (pPf) and TCf flows. The expected scaling of the turbulent fraction, $F_t = w_t/\lambda$, is obtained on the basis that the width of the turbulent bands $w_t$ is essentially constant in the vicinity of $R_{sl}$. In all cases, we find a remarkable agreement between the reported data and the scaling predicted in the proximity of a saddle-loop bifurcation (Eq.~\eqref{eq:logdiv}). Note the deviation below $R_{sl}$, as the pattern breaks down and the other critical point $R_g<R_{sl}$ for directed percolation is approached (see e.g \cite{klotz2022phase}).

Altogether, the linear instability of the homogeneous turbulent flow leads to a plethora of coexisting metastable patterns, and the fluctuations select the wavelength of the observed nonlinear pattern. The limit cycle, associated with the selected wavelength, grows, distorts, and finally hits the laminar fixed point in a global saddle-loop bifurcation, leading to a logarithmic divergence of the wavelength and, as a consequence, of the turbulent fraction. In practice, the divergence is regularized by a crossover to richer dynamics involving more than one dimension. Nevertheless, the global bifurcation and the associated singularity control the dynamics in a critical range accessible experimentally and numerically. This calls for refined simulations and experiments in the vicinity of the crossover.  

The present model shares some similarities with the two-fields one-dimensional model introduced in~\cite{Barkley-2011} to describe the transition to turbulence in pipe flow. Both models describe excitable dynamics, allowing for stable localized pulse solutions corresponding to the turbulent spots. However, while in~\cite{Barkley-2011} the inhibiting role of turbulence comes from an increased dissipation of the turbulent energy, here it arises from reduced turbulent energy production (see Supp. Mat. sec~I.D). This is actually responsible for a structurally different organization of the vector field $(f,g)$. In~\cite{Barkley-2011}, the nullclines have slopes of opposite signs when they cross, here the slopes have the same sign. This last property being a necessary condition for a Turing instability of the upper branch, \cite{Barkley-2011} does not exhibit laminar turbulent periodic patterns. 

Finally, the scenario described above is not unique to the subcritical transition to turbulence. Reaction-diffusion is most common in many other contexts including biology, chemistry or ecology \cite{turing_chemical_1952,pearson1993complex,lengyel1992chemical,rietkerk2021evasion} and generically leads to pattern formation via a linear instability, be the instability of strict Turing type or not. Making predictions beyond the weakly non-linear regime in such spatially extended systems is generally challenging. When the pattern wavelength increases away from the instability, the methodology presented here is a promising path to unveil the general scenario of a saddle-loop bifurcation~\cite{gaspard_measurement_1990,plaza1997excitability}.

{\it --- Acknowledgments ---}
PK acknowledges financial support from the French Ministry of Education and Research. We thank J.A. Sherratt for useful discussions on continuation methods. JFM thanks the support of the Competition for Research Regular Projects, year 2023, code LPR23-06, Universidad Tecnol\'ogica Metropolitana.
\vspace{-5mm}
\bibliography{ST2T.bib,other.bib,Yohann.bib}

% \begin{thebibliography}{0}
% \end{thebibliography}
%\input{SuppMat}

\end{document}